\documentclass[doublecol]{epl2} 
\usepackage{graphicx}  % needed for figures
\usepackage{dcolumn}   % needed for some tables
\usepackage{bm}        % for math
\usepackage{amssymb}   % for math
\usepackage{amsthm,amsmath,amsfonts,amssymb,hyperref}
\newtheorem{theorem}{Theorem}

\title{Calculation of the critical temperature of a dilute Bose gas in the Bogoliubov approximation}
\shorttitle{Calculation of the Critical Temperature of a Dilute Bose Gas} %Insert here a short version of the title if it exceeds 70 characters

\author{M. Napi\'orkowski\inst{1,2} \and R. Reuvers\inst{3,4} \and J.P. Solovej\inst{3}}
\shortauthor{M. Napi\'orkowski \etal}

\institute{                    
  \inst{1} Institute of Science and Technology Austria, Am Campus 1, 3400 Klosterneuburg, Austria\\
  \inst{2} Department of Mathematical Methods in Physics, Faculty of Physics, University of Warsaw, Pasteura 5, 02-093 Warsaw, Poland\\
  \inst{3} QMATH, Department of Mathematical Sciences, University of Copenhagen, Universitetsparken 5, DK-2100 Copenhagen \O, Denmark\\
  \inst{4} DAMTP, Centre for Mathematical Sciences, University of Cambridge, Wilberforce Road, Cambridge CB3 0WA, United Kingdom}


\abstract{Following an earlier calculation in 3D, we calculate the 2D critical temperature of a dilute, translation-invariant Bose gas using a variational formulation of the Bogoliubov approximation introduced by Critchley and Solomon in 1976. This provides the first analytical calculation of the Kosterlitz--Thouless transition temperature that includes the constant in the logarithm.}

\begin{document}

\maketitle

\section{Introduction}
Despite experimental realizations of cold-atom Bose--Einstein condensation (BEC) in 1995, it remains difficult to measure the critical temperature change caused by interactions \cite{Smith}. This is especially true for homogeneous gases, for which, in spite of experiments in trapped gases \cite{Smith2} and novel experimental techniques \cite{GauntSmith}, no measurements were ever made.

To theorists, homogeneous gases also pose challenges: a \textit{non-interacting}, or \textit{free}, 3D Bose gas with density $\rho$ forms a BEC below
\begin{equation}
\label{Tfc}
T_{\rm fc}=4\pi\zeta(3/2)^{-2/3}\rho^{2/3}
\end{equation}
(with $\hbar=2m=k_B=1$), but how does an interaction change this \textit{free critical temperature}? Feynman \cite{Feynman-53} used path integrals to qualitatively answer this question for liquid helium. He predicted that the potential increases the effective mass, lowering the critical temperature---something that had already been measured.

To make quantitative predictions, various simplifications were considered. For a hard-core Bose gas with core radius (or scattering length) $a$, one can assume that the gas is \textit{dilute} and apply perturbation theory. The relevant parameter is $\rho^{1/3}a\ll1$, implying that the particles tend to be far apart on the scale of the interaction.

Studying this set-up, Lee and Yang \cite{LeeYan-58} replaced the hard-core potential with a pseudopotential \cite{HuaYan-57,LeeHuaYan-57} and simplified the resulting Hamiltonian with the Bogoliubov approximation \cite{Bogoliubov-47b}. Through its energy spectrum, they found that the critical temperature shifts by an amount proportional to $\rho^{1/3}a$, that is, 
\begin{equation}
\label{exprTc}
T_{\rm{c}}=T_{\rm{fc}}(1+1.79(\rho^{1/3}a)+o(\rho^{1/3}a)),
\end{equation}
see (A2) in \cite{LeeYan-58}. This result was not taken seriously as the approach wrongly predicts a first-order phase transition.

What followed was an intense debate about the size of the critical temperature shift \cite{Andersen-04,Baymetal-01}. Motivated by the discrepancy between the observed decrease in $T_{\rm c}$ for helium and \eqref{exprTc}, some early theoretical results such as \cite{FetWal} disagreed with the predicted temperature increase. Other papers disputed the linear dependence on $\rho^{1/3}a$ (\cite{GKW,Hua1,Hua2} predict exponents of 1/2, 3/2 and 1/2, respectively). Ultimately field-theoretic methods showed that expression \eqref{exprTc} was accurate \cite{Stoof,BijSto-96,Baymetal-99,Yukalov}, and Monte Carlo simulations predicted that the constant's value 1.79 should be closer to 1.3 \cite{Arnold,Kash,NhoLan-04}. The fact that Bogoliubov's excitation spectrum manifestly predicts the correct expression seems to be little known. In our view, it is a worthwhile point to emphasize because the techniques involved are standard and accessible compared to more accurate, renormalization-based methods.

In a recent work \cite{NapReuSol2-15}, we approach the Bogoliubov approximation from a variational angle, rediscovering a model first introduced in \cite{CriSol-76} but scarcely used since. This allows us to include some interaction terms not present in the standard Bogoliubov Hamiltonian. Although the approach is still Bogoliubov-based, and so wrongly predicts a first-order phase transition, it sharpens \eqref{exprTc} to
\begin{equation}
T_{\rm{c}}=T_{\rm{fc}}(1+1.49(\rho^{1/3}a)+o(\rho^{1/3}a)).
\end{equation}
This analytical result is the closest to numerical predictions to date, which means it does better than more advanced methods, but that can of course be a coincidence.

In this paper, we repeat our calculation for a dilute 2D Bose gas to give the first analytical calculation of the Kosterlitz-Thouless transition temperature that includes the constant in the logarithm.

That may require some explanation. Our 3D calculation \cite{NapReuSol2-15} closely resembles the approach presented here (in fact, the paper can be read as an outline of that case). This could seem surprising because the phase transition in 3D involves BEC, which does not occur in homogeneous 2D Bose gases according to the Mermin--Wagner--Hohenberg theorem \cite{Hohenberg}. In reality, the 2D phase transition we study here is the formation of a quasi-condensate \cite{Popov,Kagan}, and it has been argued that Bogoliubov theory can accurately describe these \cite{MoraCastin} (see also \cite{Review2D} for a review). Quasi-condensates have been observed experimentally \cite{2Dexp}, and do not necessarily indicate the superfluid phase associated with Kosterlitz--Thouless physics \cite{KosterlitzThouless}, but as we explain before Theorem \ref{Crit}, (quasi-)condensation and superfluid pairing always occur together in this model and so our result is indeed this model's prediction of the KT transition temperature.

As in the 3D case, we expect that renormalization group approaches describe the full physics more accurately, but nevertheless the Bogoliubov Hamiltonian predicts the anticipated critical temperature. Historically, it was found that the dilute gas in 2D ($\rho^{1/2}a\ll1$) has a critical temperature \cite{Popov} in terms of $b=1/|\ln(\rho a^2)|\ll1$,
\begin{equation}
T_{\rm{c}}\approx-\frac{4\pi\rho}{\ln(b)},
\end{equation}
and this was confirmed with Bogoliubov theory by Fisher and Hohenberg \cite{FisherHohenberg} and later proved as an exact upper bound in \cite{SeiUel-09}. We should point out that both \cite{Popov} and \cite{FisherHohenberg} assume $\ln(1/b)\gg1$ instead of the appropriate and weaker $b\ll1$ \cite{Review2D, Schick}, and that \cite{FisherHohenberg} still requires renormalization group techniques. Here, we only use $b\ll1$ and the variational model to find leading behaviour 
\begin{equation}
\label{this expression}
T_{\rm{c}}=4\pi\rho\left(\frac{1}{\ln(\xi/4\pi b)}+o(1/\ln^2{b})\right)
\end{equation}
with $\xi=14.4$. Besides the numerical prediction of $\xi=380$ \cite{Prokof'ev}, no calculations of $\xi$, in particular no analytical ones, were ever done.

\section{Set-up}
We start from the Hamiltonian for a gas of $N$ bosons with a repulsive pair interaction $V$ in a $n$-dimensional box $\left[-l/2,l/2\right]^n$ and periodic boundary conditions, where $n=2,3$. In units $\hbar=2m=k_B=1$, 
\begin{equation}
\label{HN1}
H_N=\sum_{1\leq i\leq N}-\Delta_i+\sum_{1\leq i<j\leq N}V_{ij},
\end{equation}
with second-quantized form in momentum space
\begin{equation}
\label{HN}
H=\sum_p p^2 a^\dagger_p a_p+\frac{1}{2l^n}\sum_{p,q,k} \widehat{V}(k) a_{p+k}^\dagger a_{q-k}^\dagger a_q a_p.
\end{equation}
The \textit{canonical} Gibbs state at temperature $T$ and particle density $\rho=N/l^n$ can be found by minimizing
\begin{equation}
\label{tomin2}
\inf_{\omega}\big[\langle H_N\rangle_\omega-TS(\omega)\big],
\end{equation}
where $\omega$ is an $N$-boson state and $S$ is the von Neumann entropy.

The \textit{grand canonical} Gibbs state at temperature $T$ and chemical potential $\mu$ is the minimizer of 
\begin{equation}
\label{tomin}
\inf_{\omega}\big[\langle H-\mu\mathcal{N}\rangle_\omega-TS(\omega)\big],
\end{equation}
where $\omega$ is now a state on the bosonic Fock space, $\mathcal{N}$ is the particle number operator and the infimum itself is the \textit{free energy}.

Throughout this paper we are working in the \textit{thermodynamic limit} $l\to\infty$. The two quantities \eqref{tomin2} and \eqref{tomin} are then related by a Legendre transform.

We say that a system displays \textit{BEC} if the 1-particle reduced density matrix of the minimizing $\omega$ of \eqref{tomin} has an eigenvalue of order 1 \cite{PenOns}. Therefore, \textit{one needs to find minimizers of \eqref{tomin} to determine $T_{\rm c}$}.

This cannot be done exactly: the exact free energy \eqref{tomin} has only been analysed in \cite{Sei-08, Yin-10}; all other results, such as \cite{ZagBru-01}, concern approximations. We will study one such model \cite{CriSol-76}. It restricts the minimization problem \eqref{tomin} to \textit{quasi-free states}, resulting in a variational upper bound to the free energy.

There are good arguments why this upper bound is accurate. The first is that Bogoliubov's approach renders the Hamiltonian quadratic in creation and annihilation operators. Ground and Gibbs states of such Hamiltonians are quasi-free states; exactly the states considered in our minimization problem. Also, quasi-free states are good trial states for the ground state energy of Bose gases \cite{Solovej-06,ErdSchYau-08,GiuSei-09} and may therefore also be for the free energy.

As we shall soon see, expressing $\langle H-\mu\mathcal{N}\rangle_\omega-TS(\omega)$ for a general quasi-free state leads to a non-linear functional \eqref{funct}. Linearizing the functional by removing the terms quartic in creation and annihilation operators, the authors of \cite{CriSol-76} conclude that the Gibbs state coincides with that of Bogoliubov's (approximated) Hamiltonian. 
Motivated by the discovery in \cite{ErdSchYau-08} that the correct first-order energy is only found when the terms quartic in creation and annihilation operators are included, we consider the functional \textit{without} the linearization---hence including interacting terms that were ignored in the original Bogoliubov approximation---and use it to give a variational calculation of  $T_{\rm c}$.

\section{Model}
We expect the particles to form a condensate at momentum $p=0$. We therefore mimic Bogoliubov's c-number substitution (justified in \cite{LieSeiYng-05}) by using a Bogoliubov transformation to include a \textit{condensate density} $\rho_0\geq0$ in the Hamiltonian \eqref{HN}, effectively replacing $a_0\rightarrow a_0+\sqrt{l^n\rho_0}$. A minimizer with $\rho_0>0$ \textit{indicates (quasi-)condensation}, whereas $\rho_0=0$ signifies its absence. 

We evaluate the expectation value of the resulting Hamiltonian for quasi-free states only, so that we can use Wick's rule to split $\langle a_{p+k}^\dagger a_{q-k}^\dagger a_q a_p\rangle$ as
\begin{equation}
\langle a^\dagger_{p+k}a^\dagger_{q-k}\rangle\langle a_q a_p\rangle+\langle a^\dagger_{p+k}a_q\rangle\langle a^\dagger_{q-k}a_p\rangle+\langle a^\dagger_{p+k}a_p\rangle\langle a^\dagger_{q-k}a_q\rangle.
\end{equation}
Assuming translation invariance and $\langle a_p a_{-p}\rangle=\langle a^\dagger_{-p} a^\dagger_p\rangle$, the two (real-valued) functions \mbox{$\gamma(p):=\langle a^\dagger_p a_p\rangle\geq0$} and \mbox{$\alpha(p):=\langle a_p a_{-p}\rangle$}, together with the number $\rho_0$, now fully determine the expectation value in \eqref{tomin}. 
Here, $\gamma(p)$ is the \textit{density of particles with momentum $p$}, and $\alpha$ describes \textit{pairing in the system}. It is well-known that \mbox{$\alpha^2\leq\gamma(\gamma+1)$}.

%We have off-diagonal long-range order (ODLRO) if $\alpha$ is not the zero-function. It is well-known that \mbox{$\alpha^2\leq\gamma(\gamma+1)$}.

Taking the thermodynamic limit $l\to\infty$, we have now evaluated the expectation in \eqref{tomin} for \textit{Bogoliubov trial states}---quasi-free states with an added condensate---resulting in the \textit{Bogoliubov free energy functional}
\begin{equation}
\label{funct}
\begin{aligned}
\mathcal{F}&_{\mu,T} (\gamma,\alpha,\rho_0)=(2\pi)^{-n}\int p^2\gamma(p)dp-\mu\rho-TS(\gamma,\alpha)\\
&+\rho_0(2\pi)^{-n}\int\widehat{V}(p)(\gamma(p)+\alpha(p))dp+\frac{1}{2}\widehat{V}(0)\rho^2\\
&+\frac{1}{2}(2\pi)^{-2n}\int\gamma(p)(\widehat{V}\ast\gamma)(p)+\alpha(p)(\widehat{V}\ast\alpha)(p)dp,
\end{aligned}
\end{equation}
with chemical potential $\mu\in\mathbb{R}$ and density \mbox{$\rho=\rho_0+\rho_\gamma$} (i.e.\ the sum of the condensate density $\rho_0$ and the density of particles with positive momentum \mbox{$\rho_\gamma=(2\pi)^{-n}\int\gamma$}). 
The entropy, defined in terms of $\beta(p)=\sqrt{(\gamma(p)+\frac{1}{2})^2-\alpha(p)^2}$, is
\begin{equation}
\begin{aligned}
S(\gamma,\alpha)&=(2\pi)^{-n}\int\left(\beta(p)+\frac{1}{2}\right)\ln\left(\beta(p)+\frac{1}{2}\right)\\
&\hspace{2cm}-\left(\beta(p)-\frac{1}{2}\right)\ln\left(\beta(p)-\frac{1}{2}\right)dp.
\end{aligned}
\end{equation}
A precise derivation can be found in the appendix of \cite{NapReuSol1-15}.

For a canonical formulation with fixed \textit{average} density $\rho$ and temperature $T$, we consider
\begin{equation}
\label{canproblem}
\mathcal{F}^{\rm can}_{\rho,T}(\gamma,\alpha,\rho_0)=\mathcal{F}_{\mu,T} (\gamma,\alpha,\rho_0)+\mu\rho
\end{equation}
for states $(\gamma,\alpha,\rho_0)$ with $\rho_0+\rho_\gamma=\rho$. This amounts to evaluating the expectation in \eqref{tomin2} for Bogoliubov trial states.

In what follows, we drop the subscripts of $\mathcal{F}^{\rm can}_{\rho,T}$ and $\mathcal{F}_{\mu,T}$. Note that in contrast to \eqref{tomin2} and \eqref{tomin}, the infima of these functionals are not automatically related by a Legendre transform because of the restricted minimization. For example, the canonical infimum is not convex in 3D \cite{NapReuSol2-15}.

\section{Results}
We now restrict to 2D and assume that the two-body interaction potential is repulsive, integrable and bounded. Its Fourier transform $\widehat{V}$ is assumed to be positive and it has its maximum at zero since $V\geq0$.

The gas is dilute, so $\rho^{1/2}a\ll1$ in 2D, which implies $b=1/|\ln(\rho a^2)|\ll1$ for the expansion parameter. To find $T_{\rm c}$, we consider temperatures that satisfy $\sqrt{T}a\ll1$; otherwise the first line in \eqref{funct} dominates and it is easy to show that $\rho_0=0$.
We also use
\begin{equation}
\label{expand}
\widehat{V}(p)=\widehat{V}(0)+Ca^2p^2+o(a^2p^2),
\end{equation}
where the first derivative is absent since $\widehat{V}$ has its maximum at zero, and the second derivative is assumed to be of order $a^2$ (in accordance with its units).\footnote{In 3D, $\rho^{1/3}a\ll1$ and $\widehat{V}(0)=O(a)$, modifying \eqref{expand} accordingly.}

\begin{figure}
\includegraphics[scale=0.28]{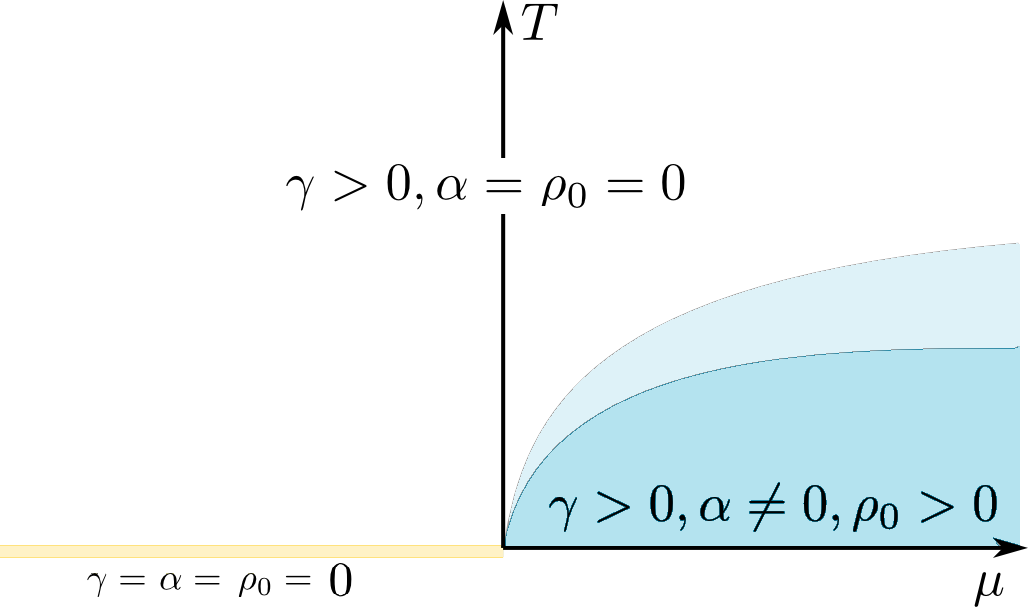}
\caption{The grand canonical phase diagram of the model. No diluteness is assumed. At $\mu\leq0$ and $T=0$, all quantities are zero, and there is no (quasi-)condensation. Increasing $T$ does not lead to a phase transition, although $\gamma$ becomes non-zero. For $\mu>0$ fixed and $T=0$, there is (quasi-)condensation. This remains the case when $T$ increases (darkest region), eventually leading to a phase transition somewhere in the lighter region before we enter the white region where $\rho_0=0$. We can only locate the phase transition exactly for $\mu\to0$; this corresponds to the dilute limit studied in Theorem \ref{Crit}.}
\label{phasediagram}
\end{figure}

In paper \cite{NapReuSol1-15}, we prove that there exist minimizers for both the grand canonical \eqref{tomin} and canonical \eqref{tomin2} minimization problems when the minimization is restricted to Bogoliubov trial states (resulting in a minimization of the functionals \eqref{funct} and \eqref{canproblem}, respectively): the grand canonical phase diagram is shown in fig.\ \ref{phasediagram}. In particular, there is a phase transition at positive $T$ for all fixed $\mu>0$. Also note that $\alpha\neq0$ iff $\rho_0>0$, so that (quasi-)condensation and pairing always occur together. Canonically, there is a phase transition at positive $T$ for all fixed $\rho>0$.

To calculate the critical temperature, the definition of diluteness $\rho^{1/2}a\ll1$ suggests that we study the canonical functional \eqref{canproblem}.

\begin{theorem}
\label{Crit}
Consider the canonical problem \eqref{tomin2} in 2D restricted to Bogoliubov trial states with $\rho=\rho_0+\rho_\gamma$ fixed, resulting in the canonical functional \eqref{canproblem} with $n=2$. The critical temperature, defined by the properties $\rho_0> 0$ if $T>T_{\rm c}$, $\rho_0= 0$ if $0\leq T<T_{\rm c}$, is
\begin{equation}
\label{critT}
T_{\rm{c}}=4\pi\rho\left(\frac{1}{\ln(\xi/4\pi b)}+o(1/\ln^2{b})\right),
\end{equation}
with $\xi=14.4$ for $\widehat{V}(0)\approx8\pi b$.
\end{theorem} 
The proof relies on a careful expansion of the free energy.

\section{Proof of Theorem \ref{Crit}}
We fix $T$ and aim to find the critical density $\rho_{\rm c}$, which can easily be inverted to \eqref{critT}.

\textit{Outline.}\hspace{0.5cm} We could try to minimize the functional by solving the Euler--Lagrange equations of \eqref{funct}. However, the terms with convolutions give non-local contributions $\widehat{V}\ast\gamma$ and $\widehat{V}\ast\alpha$. Even with a Fourier transform, this cannot be solved. 
We therefore approximate these terms so that we obtain a functional $\mathcal{F}^{\rm sim}$ that \textit{can} be minimized explicitly in $\gamma$ and $\alpha$. We expand the resulting energy integrals, and finally minimize in $\rho_0$ to determine whether it is zero or not.

To sketch this once more,
\begin{equation}
\label{appr}
\inf_{\substack{\text{$(\gamma,\alpha,\rho_0)$}\\\text{$\rho_0+\rho_\gamma=\rho$}}}
\mathcal{F}^{\rm can}\approx\inf_{0\leq\rho_0\leq\rho}\inf_{\substack{\text{$(\gamma,\alpha)$}\\\text{$\rho_\gamma=\rho-\rho_0$}}}\mathcal{F}^{\rm sim},
\end{equation}
where the infimum over $\gamma$ and $\alpha$ is calculated explicitly, then expanded in $b\ll1$, and finally minimized in $\rho_0$.\\

\textit{Step 1a.}\hspace{0.2cm} To approximate the convolution term involving $\gamma$, we use a comparison with the free Bose gas. Its energy is given exactly by
\begin{equation}
\mathcal{F}_{0}(\gamma)=(2\pi)^{-2}\int p^2\gamma(p)dp-TS(\gamma,0),
\end{equation}
whose minimizer for fixed $\rho$ is
\begin{equation}
\label{sommin}
\gamma_{\mu(\rho)}(p)=\frac{1}{e^{(p^2-\mu(\rho))/T}-1},
\end{equation}
where $\mu(\rho)\leq0$ is such that $(2\pi)^{-2}\int\gamma_{\mu(\rho)}=\rho$. Since the integral diverges as $\mu(\rho)\to0$, this definition works for all $\rho\geq0$.

We would like to show that the minimizing $\gamma$ for the interacting problem lives on the same scale as $\gamma_{\mu(\rho)}$, that is, most particles have momentum $|p|\leq O(\sqrt{T})$. Indeed, a careful comparison shows that the minimizer has to satisfy
\begin{equation}
\label{step1}
\mathcal{F}_0(\gamma_{\mu(\rho_\gamma)})\leq \mathcal{F}_0(\gamma)\leq \mathcal{F}_0(\gamma_{\mu(\rho)})+\rho^2\widehat{V}(0),
\end{equation}
which says that the energy does not deviate much from the minimal energy in the free case. Because of the $\int p^2\gamma$ term, this means that $\gamma$ cannot be very large for \mbox{$|p|\gg\sqrt{T}$}.

For $|p|\leq O(\sqrt{T})$, \eqref{expand} implies that \mbox{$|\widehat{V}(p)-\widehat{V}(0)|=O(Ta^2)\ll 1$}.
Since $\gamma$ is only large on $|p|\leq O(\sqrt{T})$, we can approximate $(2\pi)^{-2}\widehat{V}\ast\gamma\approx\widehat{V}(0)\rho_\gamma$. 

To be more precise: a careful analysis shows
\begin{equation}
\left|(2\pi)^{-2}\rho_0\int\widehat{V}(p)\gamma(p)dp
- \rho_0\widehat{V}(0)\rho_\gamma\right|= O(\rho^2(\rho^{1/2}a)),
\end{equation}
and similar estimates for the term involving the convolution. This error is negligible compared to contributions of order $T^2b$---the order to which we shall expand the energy---for example if $\rho\leq O(T/\sqrt{b})$. The latter we can assume without loss of generality by showing that $\mathcal{F}^{\rm can}(\gamma,0,0)$ grows rapidly compared to $\mathcal{F}^{\rm can}(0,0,\rho_\gamma)$, demonstrating that $\rho_0>0$ if the density is large enough.\\

\textit{Step 1b.}\hspace{0.2cm} The strategy for the convolution term with $\alpha$ is different.
Adapting ideas in \cite{ErdSchYau-08} to the 2D case, we expect $\alpha$ to be related to the function $w:=2bw_0$, where $w_0$ is the \textit{scattering solution} satisfying
\begin{equation}
-\Delta w_0+\frac12 Vw_0=0,
\end{equation}
with $w_0(x)\sim \ln(|x|/a)$ as $|x|\to\infty$.

To work towards a good approximation for $\alpha$, we define
\begin{equation}
\label{alpha0}
\alpha_0:=(\rho_0+t_0)\widehat{w}-(2\pi)^2 \rho_0\delta_0,
\end{equation}
where $\delta_0$ is a delta function and $-\rho_0\leq t_0\leq0$ is an additional parameter that will be tuned to achieve a self-consistency equation \mbox{$\int(\alpha-\alpha_0)=0$}.
In 2D, $w$ has logarithmic asymptotic behaviour, and its Fourier transform is more complicated than in the 3D case. Computing it, we find
\begin{equation}
\label{alpha02}
\alpha_0=(2\pi)^2 t_0\delta_0-(\rho_0+t_0)\left(\frac{\widehat{Vw}(p)}{2p^2}\chi_{|p|> p_0}-\langle h,.\rangle\right),
\end{equation}
where $\chi$ indicates an indicator function and $p_0$ is a momentum scale $2\rho^{1/2}e^{-\Gamma}$ involving the Euler--Mascheroni constant $\Gamma$.
The third contribution in \eqref{alpha02} is a distribution that acts on test functions $f$ as
\begin{equation}
\langle h,f\rangle=\int_{|p|\leq p_0}\frac{\widehat{Vw}(p)f(p)-\widehat{Vw}(0)f(0)}{2p^2}.
\end{equation}

The motivation for defining $\alpha_0$ in this way is as follows: at momentum scales bigger than $\sqrt{Tb}$, we expect $\alpha$ to by related to the scattering solution. Its structure on smaller scales is more complicated, but the exact shape is irrelevant. We approximate this part by a $\delta$-function and eventually optimize our approximation in $t_0$. 

So how does the guess \eqref{alpha0} help? We add and subtract terms to replace the convolution term with $\alpha$ by 
\begin{equation}
\int(\alpha-\alpha_0)(p)(\widehat{V}\ast(\alpha-\alpha_0))(p)dp,
\end{equation}
which we later show to be small for the minimizing $\alpha$. 
By doing this we have of course introduced terms involving $\widehat{V}\ast\alpha_0$, but
\begin{equation}
(2\pi)^{-2}\widehat{V}*\alpha_0(p)=(\rho_0+t_0)\widehat{Vw}(p)-\rho_0\widehat{V}(p),
\end{equation}
so that no convolution terms remain in our functional. This has the added effect that $\widehat{V}$ gets replaced by $\widehat{Vw}$ in the term linear in $\alpha$, but this Fourier transform is well-defined and satisfies $\widehat{Vw}(0)=8\pi b$. To simplify the resulting functional, we make sure to obtain a similar replacement for the $\rho_0\int\widehat{V}\gamma$-term.\\

\textit{Step 1c.}\hspace{0.2cm} To specify \eqref{appr}, we define
\begin{equation}
\label{errors}
\begin{aligned}
E_1&:=\frac12(2\pi)^{-4}\int(\alpha-\alpha_0)(p)(\widehat{V}\ast(\alpha-\alpha_0))(p)dp\\
E_2&:=\rho_0\left((2\pi)^{-2}\int\widehat{V}(p)\gamma(p)dp-\widehat{V}(0)\rho_\gamma\right)\\
E_3&:=-(\rho_0+t_0)\left((2\pi)^{-2}\int\widehat{Vw}(p)\gamma(p)dp-\widehat{Vw}(0)\rho_\gamma\right)\\
E_4&:=\frac12(2\pi)^{-4}\int\gamma(p)(\widehat{V}\ast\gamma)(p)dp-\frac12\widehat{V}(0)\rho_\gamma^2,
\end{aligned}
\end{equation}
and following Steps 1a and 1b, we now derive a simplified functional $\mathcal{F}^{\rm sim}$ satisfying\
\begin{equation}
\label{bounds}
\mathcal{F}^{\rm{can}}(\gamma,\alpha,\rho_{0})-\mathcal{F}^{\rm{sim}}(\gamma,\alpha,\rho_0)=E_1+E_2+E_3+E_4.
\end{equation}

\textit{Step 2a.}\hspace{0.2cm} Now that we have $\mathcal{F}^{\rm sim}$, we calculate and expand its minimum as a function of $\rho$ and $\rho_0$. It turns out that we need to expand to order $T^2b$ to derive \eqref{critT}.

The part of $\mathcal{F}^{\rm sim}$ that depends on $\gamma$ and $\alpha$ is
\begin{equation}
\label{simplestf}
\begin{aligned}
(2\pi)^{-2}\left[\right.\int p^2\gamma(p)dp+(\rho_0+t_0)\int\widehat{Vw}(p)(\gamma(p)+\alpha(p))dp\\
+(\rho_0+t_0)^2\int\frac{\widehat{Vw}(p)^2-\chi_{|p|\leq p_0}\widehat{Vw}(0)^2}{4p^2}dp\left.\right]-TS(\gamma,\alpha).
\end{aligned}
\end{equation}
Given $\rho$ and $\rho_0$ we find minimizers $\gamma^{\rho_0,\rho}$ and $\alpha^{\rho_0,\rho}$ by adding a Lagrange multiplier term $\delta\rho_\gamma$ (with $\delta\geq0$) to this expression.
The dominant contribution of \eqref{simplestf} to the energy $\mathcal{F}^{\rm sim}(\gamma^{\rho_0,\rho},\alpha^{\rho_0,\rho},\rho_0)$ is
\begin{equation}
\label{enexample}
(2\pi)^{-2}T\int\ln(1-e^{-T^{-1}\sqrt{(p^2+\delta)^2+2(p^2+\delta)(\rho_0+t_0)\widehat{Vw}(p)}})dp,
\end{equation}
which resembles the energy of the free gas. \\

\textit{Step 2b.}\hspace{0.2cm} We now expand \eqref{enexample}. To do this, we judiciously define
\begin{equation}
\label{replacements}
\rho_0=\frac{\sigma}{8\pi}T, \hspace{0.3cm} \rho=\frac{\ln(k)-\ln(b)}{4\pi}T,  \hspace{0.3cm} \delta=dbT, \hspace{0.3cm} t_0=\frac{\tau}{\sigma}\rho_0,
\end{equation}
where $\sigma,k,d,\tau$ are parameters of order 1. 

Changing variables $p\to \sqrt{T}p$ and using \eqref{expand} for $\widehat{Vw}$, we expand \eqref{enexample} for $b\ll1$, resulting in
\begin{equation}
\begin{aligned}
&\frac{T^2}{4\pi}\left[\right.-\frac16\pi^2-b\ln(b)(d+\sigma+\tau)+b\left(\right.d+\sigma+\tau\\
&-\frac12(d+2\sigma+2\tau)\ln(d+2\sigma+2\tau)-\frac12d\ln(d)\left.\right)+o(b)\left.\right].
\end{aligned}
\end{equation}

Of course, there is a relation between $d$, $k$ and $\sigma$, since $d$ was a Lagrange multiplier. 
We eliminate $d$ by calculating, expanding and solving \mbox{$\int\gamma^{\rho_0,\rho}=\rho-\rho_0$}, finding
\begin{equation}
\label{d}
d=-(\sigma+\tau)+\sqrt{(\sigma+\tau)^2+e^\sigma/k^2}.
\end{equation}

\textit{Step 2c.}\hspace{0.2cm} We now use \eqref{bounds} to show that we can accurately approximate the energy. We claim the minimizers have to satisfy
\begin{equation}
|\mathcal{F}^{\rm can}(\gamma,\alpha,\rho_0)-\mathcal{F}^{\rm sim}(\gamma^{\rho_0,\rho},\alpha^{\rho_0,\rho},\rho_0)|=o(T^2b),
\end{equation}
There are two bounds to show. 

The lower bound follows from \eqref{bounds} and the a priori results from Step 1a. Note that no a priori information on $\alpha$ is needed since $E_1\geq0$.

For the upper bound, we simply verify that all errors in \eqref{errors} are $o(T^2b)$ for $\gamma^{\rho_0,\rho}$ and $\alpha^{\rho_0,\rho}$.
This is only non-trivial for $E_1$, for which it suffices to choose $\tau$ such that \mbox{$\int(\alpha^{\rho_0,\rho}-\alpha_0)=0$}. We conclude
\begin{equation}
\label{tau}
\tau=\ln\left(\frac{d}{d+2\sigma+2\tau}\right)+o(1),
\end{equation}
which can be used to eliminate $\tau$.\\

\begin{figure}
\includegraphics[scale=0.42]{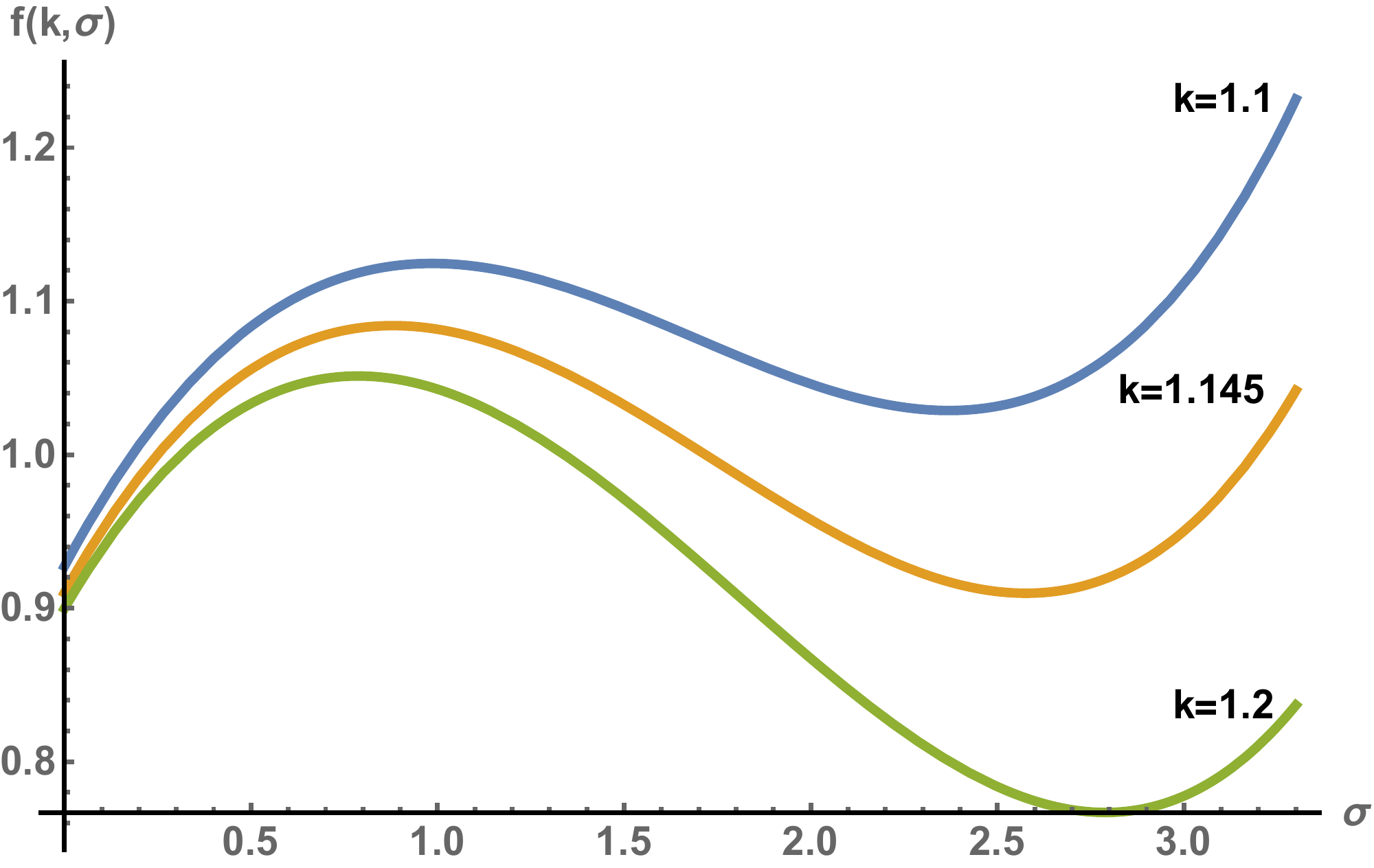}
\caption{Plots of the part of the free energy $f(k,\sigma)$ between the square brackets in \eqref{f} for three values of $k$, where $\rho=\ln(k/b)T/4\pi$ and $\rho_0=\sigma T/8\pi$. For $k=1.1$, $\sigma=\rho_0=0$ gives the lowest energy: no quasi-condensation. For $k=1.2$, the minimum occurs at some $\rho_0>0$: quasi-condensation. The critical value is $k_c=1.145$.}
\label{123}
\end{figure}

\textit{Step 3.}\hspace{0.2cm} We have now reduced \eqref{appr} to
\begin{equation}
\inf_{\substack{\text{$(\gamma,\alpha,\rho_0)$}\\\text{$\rho_0+\rho_\gamma=\rho$}}}\mathcal{F}^{\rm can}=\frac{T^2}{4\pi}\cdot\inf_{\sigma\geq0}f(k,\sigma)+o(T^2b),
\end{equation}
with, in the specific case $\widehat{V}(0)\to8\pi b$,
\begin{equation}
\label{f}
\begin{aligned}
f&(k,\sigma)=-\frac16\pi^2+2b\ln^2b-4b\ln(b)\ln(k)\\
&+b\left[\right.d+\sigma+\tau-(\sigma+\tau)\ln(d+2\sigma+2\tau)\\
&+2\ln^2k+\frac14(\sigma+\tau)^2-\frac12\tau^2-(\sigma+\tau)\ln(k)\left.\right] ,
\end{aligned} 
\end{equation}
and $d$ and $\tau$ as in \eqref{d} and \eqref{tau}. We can determine $\rho_{\rm c}$ by fixing $k$ and checking when the minimizing $\sigma$ changes from zero to non-zero, as illustrated in fig.\ \ref{123}. We find $\rho_{\rm c}=\ln(1.145/b)T/4\pi$, which can be rewritten as the desired $T_{\rm c}$ \eqref{critT}. Note that this strategy works for any value of $\widehat{V}(0)$, not just $8\pi b$, but the expression \eqref{f} simplifies in this case.

\section{Conclusion and discussion}
Adapting our earlier calculation in 3D, we analytically calculate the Kosterlitz-Thouless transition temperature of a dilute, translation-invariant Bose gas in 2D, finding \eqref{this expression} with $\xi=14.4$.

We use a variational model that can be seen as a reformulation of Bogoliubov's approximation, although, in fact, it is slightly more accurate. The relevant trial states are quasi-free states with an added (quasi-)condensate. As any approach that relies on Bogoliubov theory, this produces unphysical results like an incorrect first-order phase transition to BEC in 3D, but, perhaps surprisingly, Bogoliubov's approximation does make accurate predictions for both the 2D and 3D critical temperatures.

To improve the model one would have to extend the class of trial states, which is both very interesting and very challenging. Another compelling direction for future research is to ask what role superfluidity plays in this model, and what we can learn from it.

\acknowledgments
We thank Robert Seiringer and Daniel Ueltschi for bringing the issue of the change in critical temperature to our attention. We also thank the Erwin Schr\"odinger Institute (all authors) and the Department of Mathematics, University of Copenhagen (MN) for the hospitality during the period this work was carried out. We gratefully acknowledge the financial support by the European Union’s Seventh Framework Programme under the ERC Grant Agreement Nos. 321029 (JPS and RR) and 337603 (RR) as well as support by the VILLUM FONDEN via the QMATH Centre of Excellence (Grant No. 10059) (JPS and RR), by the National Science Center (NCN) under grant No. 2016/21/D/ST1/02430 and the Austrian Science Fund (FWF) through project Nr. P 27533-N27 (MN).


\begin{thebibliography}{0}
\bibitem{Andersen-04}
  \Name{Andersen J.O.}
  \REVIEW{Rev. Mod. Phys.}{76}{2004}{599}.

\bibitem{Arnold}
  \Name{Arnold P. \and Moore G.}
  \REVIEW{Phys. Rev. Lett.}{87}{2001}{120401}.

\bibitem{Baymetal-99}
  \Name{Baym G., Blaizot J.-P., Holzmann M., Lalo\"e F. \and Vautherin D.} 
  \REVIEW{Phys. Rev. Lett.}{83}{1999}{1703}.

\bibitem{Baymetal-01}
  \Name{Baym G., Blaizot J.-P., Holzmann M., Lalo\"e F. \and Vautherin D.}
  \REVIEW{Eur. Phys. J. B}{24}{2001}{107--124}.

\bibitem{BijSto-96}
  \Name{Bijlsma M. \and Stoof H.T.C.}
  \REVIEW{Phys. Rev. A}{54}{1996}{5085}.

\bibitem{Bogoliubov-47b}
  \Name{Bogoliubov N.N.}
  \REVIEW{J. Phys. (USSR)}{11}{1947}{23}.

\bibitem{2Dexp}
\Name{Clade P., Ryu C., Ramanathan A., Helmerson K. \and Phillips W.D.}
\REVIEW{Phys. Rev. Lett.}{102}{2009}{170401}.

\bibitem{CriSol-76}
  \Name{Critchley R.H. \and Solomon A.}
  \REVIEW{J. Stat. Phys.}{14}{1976}{381--393}.

\bibitem{ErdSchYau-08}
  \Name{Erd\H{o}s L., Schlein B. \and Yau H.-T.}
  \REVIEW{Phys. Rev. A}{78}{2008}{053627}.

\bibitem{FetWal}
  \Name{Fetter A.L. \and Walecka J.D.}
  \Book{Quantum theory of many-particle systems}
  \Publ{Courier Corporation}
  \Year{2003}.

\bibitem{Feynman-53}
  \Name{Feynman R.P.}
  \REVIEW{Phys. Rev.}{91}{1953}{1291}.

\bibitem{FisherHohenberg}
  \Name{Fisher D.S. \and Hohenberg P.C.}
  \REVIEW{Phys. Rev. B}{37}{1988}{4936}.

\bibitem{GauntSmith}
  \Name{Gaunt A.L., Schmidutz T.F., Gotlibovych I., Smith R.P. \and Hadzibabic Z.}
  \REVIEW{Phys. Rev. Lett.}{110}{2013}{200406}.

\bibitem{GiuSei-09}
  \Name{Giuliani A. \and Seiringer R.}
  \REVIEW{J. Stat. Phys.}{135}{2009}{915--934}.

\bibitem{GKW}
  \Name{Glassgold A.E., Kaufman A.N. \and Watson K.M.}
  \REVIEW{Phys. Rev.}{120}{1960}{660}.

\bibitem{Hohenberg}
  \Name{Hohenberg P.C.}
  \REVIEW{Phys. Rev.}{158}{1967}{383}.

\bibitem{Hua1}
  \Name{Huang K.}
  \Book{Studies in Statistical Mechanics}
  \Editor{J. deBoer \and G. Uhlenbeck}
  \Vol{II}
  \Publ{North-Holland}
  \Year{1964}.

\bibitem{Hua2}
  \Name{Huang K.}
  \REVIEW{Phys. Rev. Lett.}{83}{1999}{3770}.

\bibitem{HuaYan-57}
  \Name{Huang K. \and Yang C.N.}
  \REVIEW{Phys. Rev.}{105}{1957}{767}.

\bibitem{Kagan}
\Name{Kagan Y., Svistunov B.V. \and Shlyapnikov G.V.}
\REVIEW{Sov. Phys. JETP}{66}{1987}{314}.

\bibitem{Kash}
  \Name{Kashurnikov V.A., Prokof'ev N.V. \and Svistunov B.V.}
  \REVIEW{Phys. Rev. Lett.}{87}{2001}{120402}.

\bibitem{KosterlitzThouless}
  \Name{Kosterlitz J.M. \and Thouless D.J.}
  \REVIEW{J. Phys. C}{6}{1973}{1181}.

\bibitem{LeeHuaYan-57}
  \Name{Lee T.D., Huang K. \and Yang C.N.}
  \REVIEW{Phys. Rev.}{106}{1957}{1135}.

\bibitem{LeeYan-58}
  \Name{Lee T. \and Yang C.N.}
  \REVIEW{Phys. Rev.}{112}{1958}{1419}.

\bibitem{LieSeiYng-05}
  \Name{Lieb E.H., Seiringer R. \and Yngvason J.}
  \REVIEW{Phys. Rev. Lett.}{94}{2005}{080401}.

\bibitem{MoraCastin}
  \Name{Mora C. \and Castin Y.}
  \REVIEW{Phys. Rev. A}{67}{2003}{053615}.

\bibitem{NapReuSol1-15}
  \Name{Napi\'orkowski M., Reuvers R. \and Solovej J.P.}
  \REVIEW{Arch. Ration. Mech. Anal.}{}{in press, 2018}{ArXiv:1511.05935}.

\bibitem{NapReuSol2-15}
  \Name{Napi\'orkowski M., Reuvers R. \and Solovej J.P.}
  \REVIEW{Commun. Math. Phys.}{}{in press, 2017}{https://doi.org/10.1007/s00220-017-3064-x}.

\bibitem{NhoLan-04}
  \Name{Nho K. \and Landau D.P.}
  \REVIEW{Phys. Rev. A}{70}{2004}{053614}.

\bibitem{PenOns}
  \Name{Penrose O. \and Onsager L.}
  \REVIEW{Phys. Rev.}{104}{1956}{576}.

\bibitem{Popov}
  \Name{Popov V.N.}
  \Book{Functional Integrals in Quantum Field Theory and Statistical Physics}
  \Publ{Reidel, Dordrecht}
  \Year{1983}.

\bibitem{Review2D}
\Name{Posazhennikova A.}
\REVIEW{Rev. Mod. Phys.}{78}{2006}{1111}.

\bibitem{Prokof'ev}
  \Name{Prokof'ev N., Ruebenacker O. \and Svistunov B.}
  \REVIEW{Phys. Rev. Lett.}{87}{2001}{270402}.

\bibitem{Schick}
\Name{Schick M.}
\REVIEW{Phys. Rev. A}{3}{1971}{1067}.

\bibitem{Sei-08}
  \Name{Seiringer R.}
  \REVIEW{Commun. Math. Phys.}{279}{2008}{595--636}.

\bibitem{SeiUel-09}
  \Name{Seiringer R. \and Ueltschi D.}
  \REVIEW{Phys. Rev. B}{80}{2009}{014502}. 

\bibitem{Smith}
  \Name{Smith R.P.}
  \Book{Universal Themes of Bose-Einstein Condensation}
  \Editor{N.P. Proukasis, D.W. Snoke \and P.B. Littlewood}
  \Publ{Cambridge University Press}
  \Year{2017}.

\bibitem{Smith2}
  \Name{Smith R.P., Campbell R.L.D., Tammuz N. \and Hadzibabic Z.}
  \REVIEW{Phys. Rev. Lett.}{106}{2011}{250403}.

\bibitem{Solovej-06}
  \Name{Solovej J.P.}
  \REVIEW{Commun. Math. Phys.}{266}{2006}{797--818}.

\bibitem{Stoof}
\Name{Stoof H.T.C.}
\REVIEW{Phys. Rev. A}{45}{1992}{8398}.

\bibitem{Yin-10}
  \Name{Yin J.}
  \REVIEW{J. Stat. Phys.}{141}{2010}{683--726}.

\bibitem{Yukalov}
  \Name{Yukalov V.I. \and Yukalova E.P.}
  \REVIEW{Laser Phys. Lett.}{14}{2017}{073001}.

\bibitem{ZagBru-01}
  \Name{Zagrebnov V.A. \and Bru J.-B.}
  \REVIEW{Phys. Rep.}{350}{2001}{291--434}.


\end{thebibliography}
\end{document}